**Regression with Large Language Models for Materials and Molecular Property Prediction**

**Authors:** Ryan Jacobs[1,*], Maciej P. Polak[1,*], Lane E. Schultz[1], Hamed Mahdavi[2,4], Vasant Honavar[2,3,4,5], Dane Morgan[1]

[1] Department of Materials Science and Engineering, University of Wisconsin-Madison, Madison, WI, USA.

[2] Department of Computer Science and Engineering, The Pennsylvania State University, University Park, PA, USA.

[3]College of Information Sciences and Technology, The Pennsylvania State University, University Park, PA, USA

[4]Artificial Intelligence Research Laboratory, The Pennsylvania State University, University Park, PA, USA

[5]Center for Artificial Intelligence Foundations and Scientific Applications, The Pennsylvania State University, University Park, PA, USA

*These authors contributed equally

## Abstract

We demonstrate the ability of large language models (LLMs) to perform material and molecular property regression tasks, a significant deviation from the conventional LLM use case. We benchmark the Large Language Model Meta AI (LLaMA) 3 on several molecular properties in the QM9 dataset and 28 materials properties. Only composition-based input strings are used as the model input and we fine tune on only the generative loss. We broadly find that LLaMA 3, when fine-tuned using the SMILES representation of molecules, provides useful regression results which can rival standard materials property prediction models like random forest or fully connected neural networks on the QM9 dataset. Not surprisingly, LLaMA 3 errors are 5-10× higher than those of the state-of-the-art models that were trained using far more granular representation of molecules (e.g., atom types and their coordinates) for the same task. Similarly, LLaMA 3 provides comparable, although slightly worse, accuracy relative to random forest and elemental descriptors when using just compound chemical description on our set of 28 materials properties. Interestingly, LLaMA 3 provides improved predictions compared to GPT-3.5 and GPT-4o. This work highlights the versatility of LLMs, suggesting that LLM-like generative models can potentially transcend their traditional applications to tackle complex physical phenomena, thus paving the way for future research and applications in chemistry, materials science and other scientific domains.

## 1. Introduction:

The application of large language models (LLMs) has traditionally been confined to natural language processing tasks, such as text generation, translation, and sentiment analysis. However, their ability to efficiently tokenize text into a latent space with meaningful distance measures offers the tantalizing possibility of expanding their applications far beyond these domains, including in fields such as chemistry and materials science.

Recent studies have begun to explore the applications of LLMs to problems in chemistry and materials science, e.g., materials or molecular representation learning,[1–6] material generation,[7] understanding and predicting behavior of catalysts,[8,9] chemical reaction prediction,[10] and several research works have attempted to train LLMs that solve diverse molecular property prediction tasks.[11–14] For example, Sadeghi et al.[15] successfully exploited the representation from pretrained LLMs for molecular and materials science tasks, where they explored zero/few-shot molecule classification with LLMs, generating semantically enriched explanations for SMILES and fine-tuning a small-scale language model for multiple downstream tasks. Shi et al. [10] developed a framework which combines textual information representations from LLMs with Graph Neural Networks (GNNs) to predict chemical reactions. Choudhary[16] used a fine-tuned GPT-2–based transformer (AtomGPT) with a modified regression head on descriptions of the structure generated with the ChemNLP[17] describer to predict properties from the JARVIS-DFT database,[18] and achieved promising results, with some results exceeding the best results obtained with traditional models. Finally, Jablonka et al.[19] explored fine-tuning of the Generative Pretrained Transformer (GPT)-3 model on molecular properties to answer chemical questions. Most notably, they fine-tuned GPT-3 to classify various molecular properties when asked questions in natural language and achieved impressive results. They noted that this approach is particularly effective in the low-data regime, encouraging a thorough exploration for data-rich regression in a fashion similar to typical regression models when they are optimized for accuracy.

Despite the preceding advances, it is not known whether state-of-the-art LLMs can perform regression of chemical and materials properties directly from textual inputs based only on a basic composition description, e.g., a compound stoichiometry for inorganic materials or a SMILES string for organic molecules. The present work explores this question through regression of materials and molecular properties with LLMs. The motivation for this work is that, if LLMs can be fine-tuned to perform accurate regression, they might provide either better or more convenient models than conventional approaches. In particular, the LLM would represent a general regression tool, reducing the need for time spent on identifying optimal featurization approaches and optimizing ML models for different specific problems. This work seeks to answer three related questions regarding the application of LLMs in the context of materials and molecular property prediction: (1) Can LLMs be trained to perform regression tasks of material and molecular properties from textual prompts? (2) How does the performance of such LLMs compare with other established property models that are either state-of-the-art model fits or based on standard regression methods? (3) How does the regression performance of LLMs depend on LLM type, (e.g., LLaMA 3 vs. GPT-3.5) and on the mode of input features used to fine-tune the LLM (e.g., using SMILES vs. InChI strings vs. atomic types and coordinates)?

To answer the above questions, we present an approach where an LLM is fine-tuned to perform material and molecular property regression tasks. The model receives composition- or molecular-based features as a textual prompt and generates the corresponding numerical target values, for example, formation energy, as textual output. This method is notable as it eliminates the need for development of extensive domain-specific chemical or materials knowledge used to perform featurization. This work provides an initial investigation to help answer the above questions. Regarding (1), we find that the LLM (LLaMA 3) can function as a useful regression model. We broadly find competitive performance of LLaMA 3 vs. random forest models fit on modest-sized datasets (e.g., a few hundred data points) of 28 materials properties, but with fit errors much higher than state-of-the-art deep neural networks on large materials and molecular properties databases. Regarding (2), we find that the mode of input for featurizing molecules matters modestly but with statistical significance (e.g., prediction errors can change by 15-20%), indicating that identifying the optimum approach for featurizing molecular and materials input is important for maximizing the regression accuracy of LLMs. In addition, we find that LLaMA 3 surpasses GPT-3.5 and GPT-4o and is practically easier to use given its open-source nature. This result implies the choice the LLM may have a large impact on the quality of the results.

## 2. Data and Methods:

The datasets used in this work consist of both molecular and materials properties. For molecular properties, we focus on the widely studied QM9 dataset,[20] where we predict the formation energy, highest occupied molecular orbital (HOMO), lowest unoccupied molecular orbital (LUMO) and HOMO-LUMO gap energies. We use two types of input features for fine-tuning LLMs to predict the QM9 molecular properties. The first approach is to provide a unique textual representation of the molecular structure and composition for each molecule, where we explore both the Simplified Molecular Input Line Entry System (SMILES) string and the International Chemical Identifier (InChI) string. The second approach is to explicitly provide the full list of atomic coordinates and element types for each molecule. For all QM9 fits, a test dataset of 10k molecules was randomly selected at the beginning of the study and then used to evaluate all LLM fits to QM9 properties. Separate training datasets of 1k, 10k and 110k molecules were selected from the remaining molecules not part of the test dataset. All QM9 properties were trained in units of Hartree (this is the default unit of the dataset). For analysis and plotting, all output values were converted to units of eV. All datasets used in this work are provided as part of the supporting information (see **Data and Code Availability)**.

For materials properties, 28 properties were considered, which form a diverse set of experimental and computed data and different property types (e.g., mechanical, thermodynamic, electronic, etc.). This data consists of 24 sets, chosen by us, and a 4 more sets, which were considered only for purposes of comparison to AtomGPT[16]. For the 24 sets, the dataset sizes span a large range, from only 137 data points for the perovskite thermal expansion coefficient dataset to 643,916 data points for the Open Quantum Materials Database (OQMD) formation energy dataset. The property and data types, dataset sizes and original data references are summarized in **Table *1***. For all these 24 materials property datasets, only the materials composition strings (e.g., "Al2O3") are used as input features for fine-tuning the LLMs. For the OQMD data, similar to the QM9 fits, a random subset of 10k materials were withheld at the start of the study and used as a test dataset. Training datasets were constructed using the materials not represented in the test set. For the remaining materials properties datasets, 20% of the data was randomly held out as a test set, with the remaining 80% used for training. This train/test split proportion was used in order to draw meaningful comparison with ML model fits to these properties from previous published work.

**Table 1.** Summary of molecular and materials property datasets investigated in this work. Abbreviations: HOMO = highest occupied molecular orbital. LUMO = lowest unoccupied molecular orbital, $D_{max}$ = maximum cast diameter, $R_c$ = critical cooling rate, TEC = thermal expansion coefficient, $T_c$ = superconducting critical temperature, OQMD = Open Quantum Materials Database.

| Property type | Property name | Data type | Number of data points | Data reference |
|---|---|---|---|---|
| Molecular | Formation energy | Computed | 133885 | [20] |
| | HOMO | Computed | 133885 | [20] |
| | LUMO | Computed | 133885 | [20] |
| | HOMO-LUMO gap | Computed | 133885 | [20] |
| Materials (data chosen by authors) | OQMD Formation energy | Computed | 643916 | [21,22] |
| | Bandgap | Experiment | 6031 | [23] |
| | Debye Temperature | Computed | 4896 | [24] |
| | Dielectric constant | Computed | 1056 | [25] |
| | Dilute solute diffusion | Computed | 408 | [26] |
| | Double perovskite bandgap | Computed | 1306 | [27] |
| | Elastic tensor (bulk modulus) | Computed | 1181 | [28] |
| | Exfoliation energy | Computed | 636 | [18] |
| | High entropy alloy hardness | Experiment | 370 | [29] |
| | Lithium conductivity | Experiment | 372 | [30] |
| | Metallic glass $D_{max}$ | Experiment | 998 | [31] |
| | Metallic glass $R_c$ | Experiment | 297 | [32] |

| | Oxide vacancy formation | Computed | 4914 | [33] |
|---|---|---|---|---|
| | Perovskite formation energy | Computed | 9646 | [34] |
| | Perovskite O p-band center | Computed | 2912 | [35] |
| | Perovskite stability | Computed | 2912 | [35] |
| | Perovskite TEC | Experimental | 137 | [36] |
| | Perovskite work function | Computed | 613 | [37] |
| | Phonon frequency | Computed | 1265 | [38] |
| | Piezoelectric max displacement | Computed | 941 | [39] |
| | Steel yield strength | Experiment | 312 | [40] |
| | Superconductivity $T_c$ | Experiment | 6252 | [41] |
| | Thermal conductivity | Computed | 4887 | [24] |
| | Thermal expansion | Computed | 4886 | [24] |
| Materials (Data matches that used in Ref. [16]) | Formation Energy | Computed | 55713 | [18] |
| | Bandgap (OptB88vdW) | Computed | 55713 | [18] |
| | Bandgap (mBJLDA) | Computed | 18167 | [18] |
| | Superconductivity $T_c$ (JARVIS) | Computed | 616 | [18] |

As mentioned above, we also compare the performance of our approach based on LLaMA 3 and the AtomGPT[16] method on the datasets used to evaluate AtomGPT's performance (last four datasets listed in **Table *1***). In this case, due to the similar nature of these methods (both being language model-based), we used the exact same input as AtomGPT, i.e. full textual descriptions of the structures, to allow for a direct comparison of the results.

All LLaMA 3 models were fine-tuned using the Unsloth and HuggingFace python packages. For all cases, the LLaMA 3 8B model variant in 4-bit mode with 16 Low-Rank Adaptation (LoRA)[42] parameters were used, resulting in a total of 41,943,040 trainable parameters. An example script showing the prompt used, the fine-tuning approach, and inference on test data is provided as part of the supporting information (see **Data and Code Availability**). All GPT models were fine-tuned using the OpenAI API version 1.38.0. For GPT training, training datasets consisted of conversations composed of one exchange – a user prompt consisting of the molecule representation (e.g. SMILES string) and an assistant response which was a value of the properties (e.g. formation energy). In all fine-tuning cases, the generative cross-entropy loss is minimized during training. Note that there is no simple connection between the generative cross entropy loss and the mean absolute error (MAE) on the target predictions, so it is not at all obvious that the present approach would yield effective models. We reiterate that we use cross entropy loss as the training metric as we are primarily interested in assessing the ability of out-of-the-box LLMs like LLaMA 3 to solve regression problems in a fully text-to-text manner, where inputs such as chemical compositions or SMILES strings are provided as text, and the output of a materials property, such as formation energy, is similarly provided as textual output.

# 3. Results and Discussion:

### 3.1. LLaMA 3 and GPT fine-tuning performance on molecular properties

In this section, we fine-tune LLaMA 3 on molecular properties in the QM9 dataset. **Figure *1*** shows the performance of fine-tuning LLaMA 3 on QM9 formation energies using SMILES strings as input. **Figure *1*A** contains a learning curve tracking the MAE on the test dataset of 10k molecules. We observe a large reduction in test data MAE from 3.184 eV (training on 1k molecules) to 0.749 eV (training on 10k molecules) down to our lowest error of 0.100 eV (training on 110k molecules). As shown in the inset of **Figure *1*A**, this observed reduction in MAE follows common power law scaling for deep learning models, with MAE = $A \times N_{train}^{B}$, ($N_{train}$ = number of training points), where here B=-0.737 and A = $10^{2.7505}$ = 562.99.

The blue points in **Figure *1*A** are test data errors for cases where the model was trained for fewer iterations. To understand the impact of training schedule more clearly, in **Figure *1*B** we plot the test data MAE as a function of training epochs, for the case of 110k training data points. The data in **Figure *1*B** are the blue and black points on the right end of **Figure *1*A** at 110k training points. Here, we observe a typical reduction in test data MAE as training epochs are increased. For this case of QM9 formation energy, the test data MAE reached a minimum of 0.100 eV at roughly 5 epochs (we note that training does not always result in an integer number of epochs based on the training iterations and batch size used). An additional test training to 10 epochs resulted in an increase of the test data MAE, so additional training beyond this point was not attempted. **Figure *1*C** contains a parity plot showing the LLM-predicted formation energies versus the true values for our best-fit case with an MAE of 0.100 eV. We see that LLaMA 3 can function as a very good regression model for molecular properties, with generally low errors and only a handful of outlier points. Despite this, the LLaMA 3 test data MAE still lies far above the state-of-the-art value, which, from the work of Zhang et al.[43] using their Physics-Aware Multiplex Graph Neural Network (PAMNet) model, achieved an extremely low MAE of only 5.9 meV, a factor of nearly 17× lower than that obtained with LLaMA 3, as discussed more below.

To provide a performance baseline and understand the extent that fine-tuning is aiding our LLaMA 3 model to predict QM9 formation energies, we compare the above performance of predicting QM9 formation energy to zero-shot prediction of QM9 formation energies using LLaMA 3 with no fine-tuning. For this zero-shot test, the same inference prompt containing composition-based input strings from fine-tuning was used here. Overall, of the 10k molecules in the test set, 5536 of them came back with no response (i.e., LLaMA 3 produced a blank string). Of the 4464 where a numerical response was returned, 120 of them were 0, and 12 of them were very large negative numbers that were clearly unphysical (i.e., < -100). When excluding these 12 very large negative predictions, we obtain a zero-shot MAE of 59.39 eV. Between the large number of blank responses and this very large MAE value on predictions which resulted in a numerical response, it is clear that LLaMA 3 has no significant out-of-the-box predictive ability on QM9 formation energies, at least when averaged over many values, and the fine-tuning approach used here is thus highly effective in enabling the LLM to learn molecular property relationships from the SMILES representation. It is worth noting that Rubungo et. al.[44] conducted a much more thorough evaluation of zero-shot capabilities of various models, with similar conclusions.

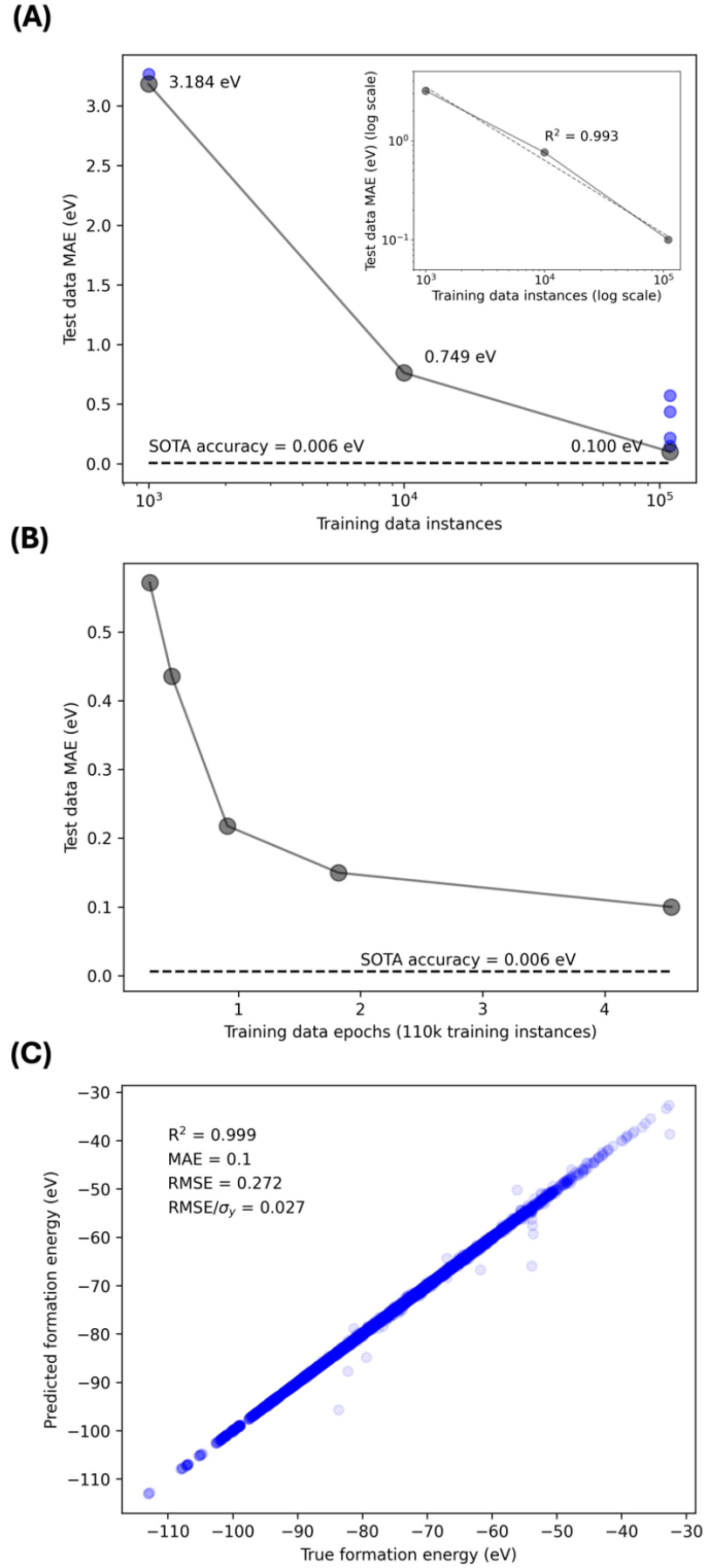

**Figure 1.** Summary of LLaMA 3 fine-tuning performance for predicting formation energy in the QM9 dataset. (A) Learning curve of test data MAE as a function of amount of training data. The inset figure depicts the linear trend of log error vs. log number of training points, showing power law scaling. The fit line has a high $R^2$ of 0.993, a slope of -0.737 and intercept of 2.7505. The blue

points indicate test MAE values for lower numbers of training iterations (i.e., non-converged runs). (B) Learning curve of test data MAE as a function of training epochs using the largest training data size of 110k points. (C) Parity plot of best LLaMA 3 fit, trained using 110k points for about 5 epochs. The values for state-of-the-art (SOTA) accuracy are from the work of Zhang et al.[43]

In **Figure 2**, we compare the performance of LLaMA 3 models trained on formation energy, HOMO level, LUMO level, and HOMO-LUMO gap energy with the PAMNet results from Zhang et al.[43] The numerical values of these results are also summarized in **Table 2**. Across these four properties, we consistently find that the LLaMA 3 model is a good regression model, but not nearly as good as state-of-the-art for predicting molecular properties, at least for the QM9 dataset. LLaMA 3 has test MAE values that are factors of 16.95×, 4.34×, 4.93×, and 4.22× larger than the PAMNet model for formation energy, HOMO level, LUMO level, and HOMO-LUMO gap, respectively, for an average ratio of 7.6×. Therefore, one can qualitatively say that LLaMA 3 is roughly 5-10× worse than state-of-the-art for predicting molecular properties in the QM9 dataset. It is perhaps not surprising that LLaMA 3 performs worse than state-of-the-art models like PAMNet, which employ extensive knowledge of molecular structure (i.e., explicit coordinates of every atom in the molecule) for training structure-aware graph neural networks, while our fine-tuning of LLaMA 3 takes as input only the SMILES string of each molecule. Therefore, part of the difference in performance may be due to PAMNet having access to detailed structural information we do not provide to the LLM. It is notable that the featurization ability of LLaMA 3 using only SMILES strings still produces a moderately low error for predicting molecular properties.

As noted above, the higher errors on QM9 properties using LLaMA 3 vs. the state-of-the-art PAMNet model may be due to the latter having access to structural information, the former being an intrinsically worse regression model, or some combination of the two. As a first step to assessing the impact of structural information on QM9 property fits, we can compare our LLaMA 3 results with fits from the work of Pinheiro et al.[45] In their work, they use the SMILES strings to construct molecular features using the RDKit and Mordred python packages, where the Mordred features are used as input to a neural network model. They trained on 100k molecules, and found test MAE values of formation energy, HOMO level, and HOMO-LUMO gap of 0.0573 eV, 0.0952 eV, and 0.1369 eV, respectively.[45] Interestingly, when compared to Pinheiro et al.[45], the formation energy MAE for our LLaMA 3 model (0.100 eV) is only a factor of 1.75× higher, the HOMO level MAE for our LLaMA 3 model (0.099 eV) is slightly higher but essentially equal, and the HOMO-LUMO gap MAE for our LLaMA 3 model (0.131 eV) is slightly lower but essentially equal. These comparisons suggest, at least for the QM9 property data, that a large portion of the performance difference between LLaMA 3 and state-of-the-art models like PAMNet is due to the way detailed structural data in used in the state-of-the-art model. While we were not able to show significant improvement in the LLaMA 3 regression using structural information (see Sec. 3.2), it is possible that performance closer to PAMNet could be obtained with LLaMA 3 through further efforts.

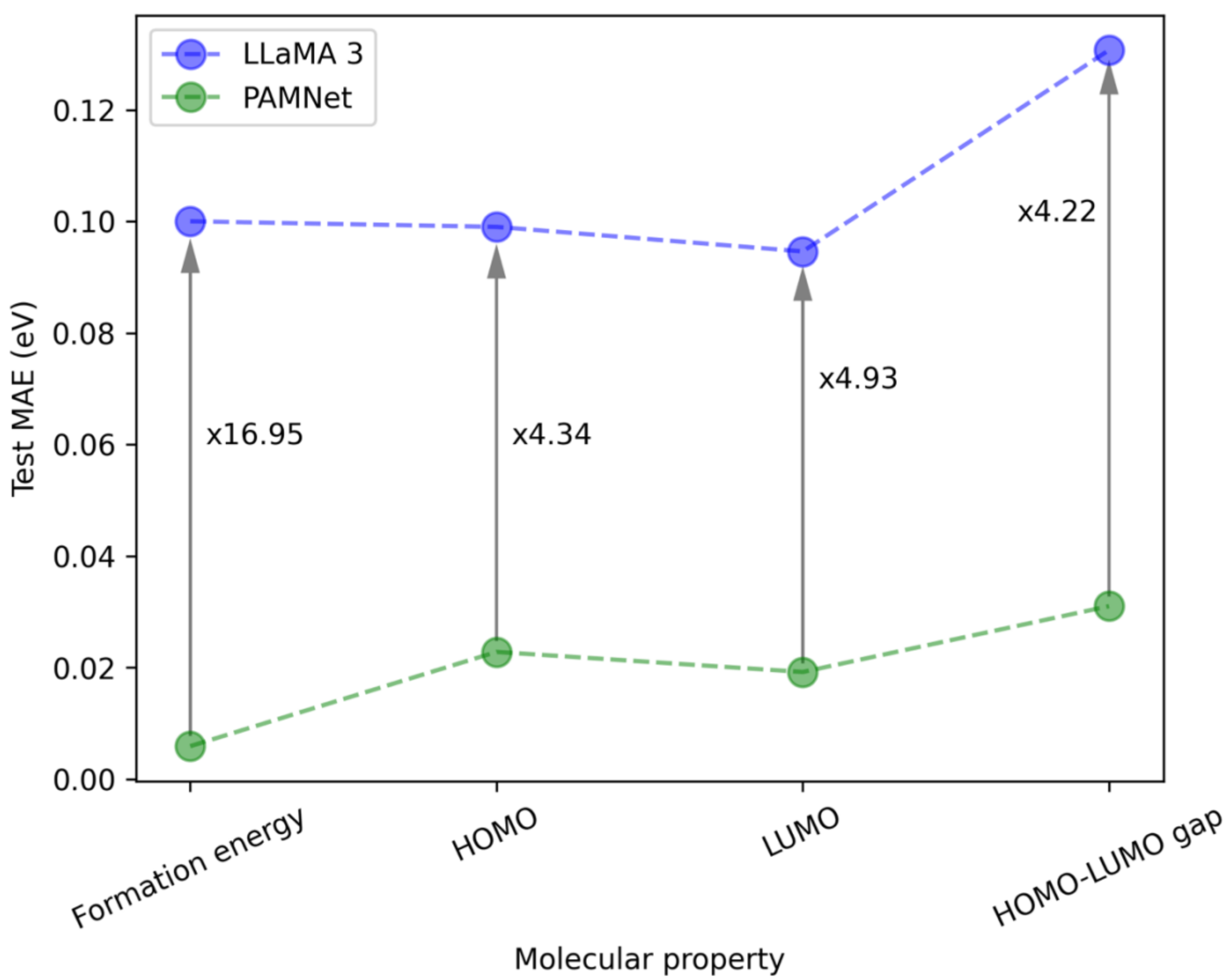

**Figure 2.** Summary of LLaMA 3 finetuning on molecular properties vs. state-of-the-art values from the work of Zhang et al.[43]

**Table 2.** Summary of LLaMA 3 molecular properties fits. For all fits, the same 110k training instances and 10k test instances were used. The MAE values for the PAMNet model are from Zhang et al.[43]

| Model | Property | Number of epochs (approx.) | LLM test MAE (eV) | PAMNet test MAE (eV) |
|---|---|---|---|---|
| LLaMA 3 | Formation energy | 5 | 0.100 | 0.0059 |
| | HOMO | | 0.099 | 0.0228 |
| | LUMO | | 0.095 | 0.0192 |
| | HOMO-LUMO gap | | 0.131 | 0.0310 |

For comparison to the LLaMA 3 results, we performed similar fine-tuning for formation energy with the OpenAI GPT family of models. Our target response in fine-tuning are floating point numbers, and GPT models tokenize numbers differently than LLaMA models. LLaMA treats each digit as a separate token, while GPT often pairs numbers together for tokenization. Assuming that treating each digit separately is advantageous, we attempted to allow such behavior when fine-tuning the GPT models by separating each digit in the target values by a space, so that GPT treats each number as a separate token. Fine-tuning OpenAI GPT models is not very flexible. At the time of this writing, these closed-source models can be fine-tuned only directly through the OpenAI API, with a very limited set of hyperparameters available for adjustment. These parameters are the number of epochs, learning rate, and batch size. We found that tuning learning rate and batch size have virtually no effect on our results, while more epochs improve the result up to around epoch 20. We fine-tuned gpt-3.5-turbo-0125 and gpt-4o-mini-2024-07-18, with virtually identical results. The best result for formation energy, obtained for gpt-4o-mini-2024-07-18, with learning rate of 1.8, batch size of 128 and 20 epochs on the same training and testing datasets as used for LLaMA 3 (8B), resulted in an MAE of 0.154 eV, about 1.5× worse than LLaMA 3 (8B). We believe that this result is not due to inherent inferiority of GPT models compared to LLaMA, but in the limited adjustability in the fine-tuning process for GPT models. OpenAI does not disclose the exact architecture or method of fine-tuning, so it is not possible to directly compare it with LLaMA using the same fine-tuning procedure. Since LLaMA models allow for more flexible fine-tuning, leading to improved results for regression, and due to the open-source character and lower cost of use, LLaMA is our model of choice for further studies.

Over the course of this work, newer LLaMA models became available. Therefore, we tested the performance of LLaMA 3.2 (8B) as well as the smaller LLaMA 3.2 (1B) on the QM9 formation energies with virtually identical results as the LLaMA 3 (8B), suggesting that the underlying model and its size may not be of key importance when it is fine-tuned for purposes that essentially break its inherent language capabilities and it is being used for regression.

### 3.2. LLaMA 3 fine-tuning performance on molecular properties: SMILES vs. InChI and explicit coordinates

In this section, we compare the use of different inputs to fine-tune LLaMA 3. We focus on the formation energy of molecules in QM9 for this analysis. In **Section 3.1**, we focused on using only the SMILES string as the input to fine-tune LLaMA 3. Here, we try fine-tuning LLaMA 3 using two different input types and compare the model performance to those trained on SMILES strings. The first input is the InChI string representation of molecules, and the second input uses explicit atomic coordinates and element types for every atom in the molecule. In all cases, we train on the same 110k training data points and predict the same 10k test data as used throughout this work.

In **Figure 3**, we compare the performance of LLaMA 3 fine-tuning for predicting QM9 properties using the SMILES vs. InChI molecule string designations as input, where all runs were trained to about 5 epochs. In all cases, we find that the use of SMILES strings resulted in lower test MAE values for all four QM9 properties investigated here. On average, the use of SMILES resulted in about 25% lower errors than using InChI strings. Given that both SMILES and InChI provide a unique representation of a particular molecule, it is interesting that the use of different molecular string designations results in statistically meaningful different fit qualities. InChI is generally longer and in some ways appears more complex than SMILES, which may be a factor. However, it is not clear whether this difference in model performance is general or some particular nuance of the ability of LLaMA 3 to featurize molecular strings and correlate them to properties of interest. We note that longer trainings beyond 5 epochs were attempted for the InChI string input, with no further reduction in error. Thus, it is not clear whether other modes of molecule featurization, for example, through the use DeepSMILES[46] or SELF-referencing Embedded Strings (SELFIES)[47] strings, a combination of different string types, or some other representation may improve the results. Identifying the best approach for encoding molecules for regression tasks to fine-tune LLMs is therefore an open research question.

Finally, we have performed an initial test of providing explicit atomic coordinates for fine-tuning LLaMA 3 for predicting QM9 formation energy. We provide this information in the format of simple XYZ files, which can be as part of the supporting information (see **Data and Code Availability**). Given the much larger amount of information contained in the explicit coordinates vs. the SMILES strings, we found training times to be longer and more training needed to attain equivalent error reduction as observed when training with SMILES. Overall, we found that training LLaMA 3 to roughly 2, 5, and 9 epochs with explicit coordinates resulted in a formation energy MAE of 155 meV, 101 meV, and 94 meV, respectively. We note that the error appears to still reduce slowly between 5-9 epochs, suggesting longer training may yield a lower error, but one which is only modestly lower than that obtained when using SMILES strings. For comparison, training on SMILES strings for 5 epochs yielded an MAE of 100 meV. Therefore, at least from this initial test, there is no substantial prediction enhancement when including explicit coordinates. We speculate that the modes in which explicit atomic coordinates (e.g., from *ab initio* calculations) provide substantial error reduction in non-LLM-based ML methods (e.g., GNNs) have not been fully realized in the context of performing regression with LLMs. Additional work needs to be done to explore the impact of more sophisticated models (e.g., LLaMA 3 70B), longer training times, and different approaches to providing structural information, or perhaps some mixture of structure and string representation, to the LLM.

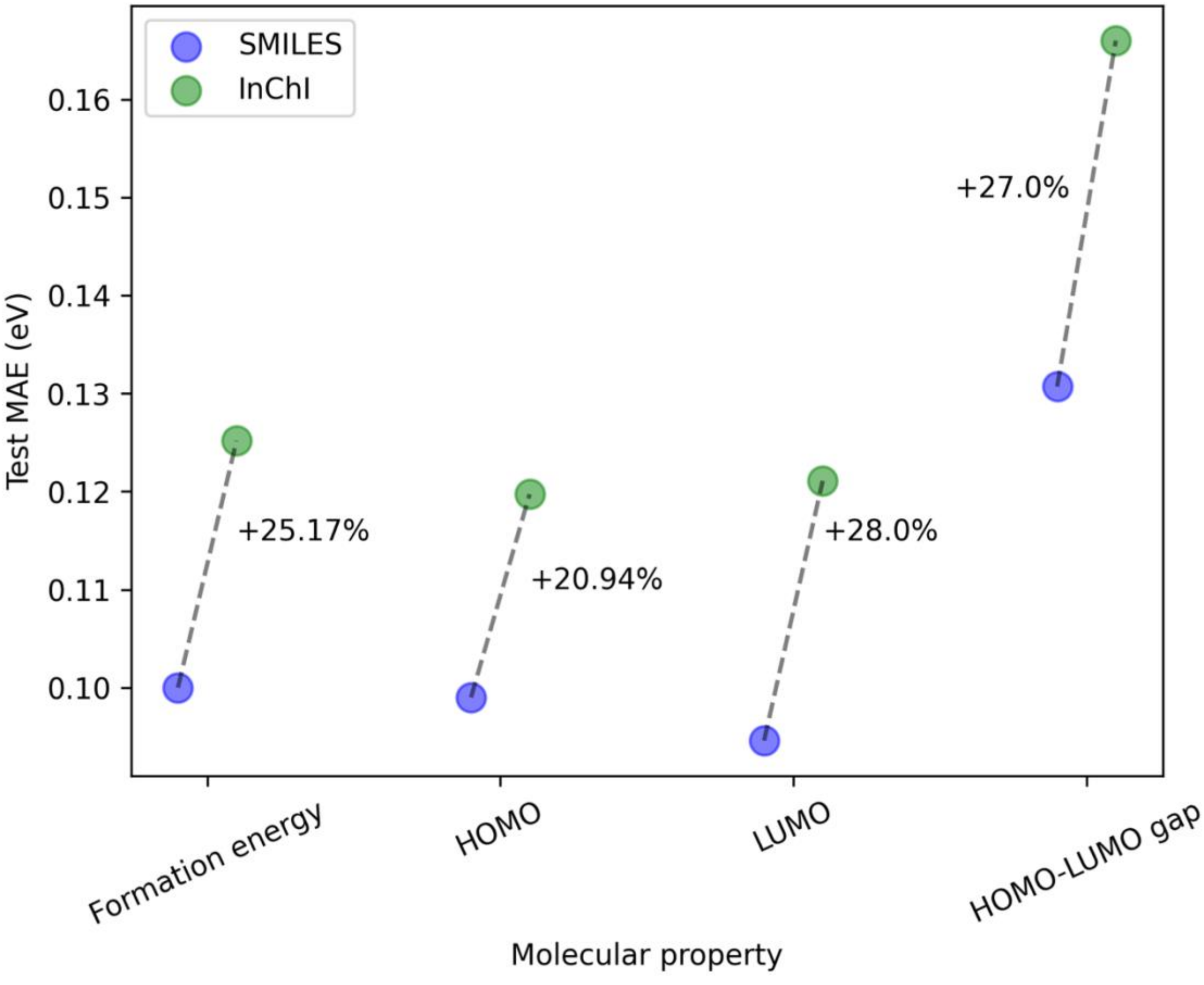

**Figure 3.** Performance of LLaMA 3 on QM9 molecular properties using SMILES vs. InChI string input. All models were trained to roughly 5 epochs.

### 3.3. LLaMA 3 fine-tuning performance on materials properties

In this section, we investigate the ability of fine-tuning LLaMA 3 for prediction of the 24 materials properties shown in **Table *1***, using only composition strings as training input. **Table 3** contains a summary of LLaMA 3 test errors on 20% left out datasets for 23 materials property datasets of small to modest size (i.e., a couple hundred to a couple thousand data points). We consider a larger 24$^{th}$ data set below. The LLaMA 3 test errors are compared with random forest models fit to a set of elemental features from the work of Jacobs et al.[48] In that work, it was found that the random forest models performed on par with other published machine learning models in the literature on the same data. Because these random forest fits are very fast, we estimate the uncertainty in the random forest MAEs by finding the standard deviation of 25 leave out 20% splits, generated from 5 sets of 5-fold cross validation. Due to the computational expense of fine-tuning LLaMA 3, only a single 20% test set was used to produce the LLaMA 3 MAE value for each property. By comparing the MAE values in **Table 3** between random forest and LLaMA 3, we find that LLaMA 3 performs better than random forest for 8 properties, LLaMA 3 and random forest perform equally well for 7 properties, and random forest performs better for 8 properties. For this comparison, LLaMA 3 was considered to perform equally with random forest if the test data MAE was within the cross-validation error bar provided in **Table 3**, and better (worse) if the LLaMA 3 MAE was lower (higher) than this value. This designation of LLaMA being better or worse than random forest is based on the assumption that the LLaMA 3 MAE value has a similar uncertainty as the standard deviation of random forest cross validation values.

**Table 3.** Summary of LLaMA 3 materials property fits compared to previously published random forest models fit using elemental features. The "±" values for random forest MAE were obtained from the standard deviation of 25 splits of leave out 20% from 5 sets of 5-fold cross validation.

| Property name | RF MAE | RF RMSE/σ | LLaMA 3 MAE | LLaMA RMSE/σ | Units | Which is better? |
|---|---|---|---|---|---|---|
| Bandgap | 0.328 ± 0.016 | 0.419 ± 0.031 | 0.274 | 0.416 | eV | LLM |
| Debye Temperature | 42.353 ± 1.571 | 0.371 ± 0.027 | 49.326 | 0.473 | K | RF |
| Dielectric constant | 0.113 ± 0.008 | 0.61 ± 0.038 | 0.097 | 0.563 | n/a (log scale) | LLM |
| Dilute solute diffusion | 0.177 ± 0.019 | 0.184 ± 0.031 | 0.202 | 0.215 | eV | RF |
| Double perovskite bandgap | 0.278 ± 0.020 | 0.254 ± 0.021 | 0.263 | 0.226 | eV | Equal |
| Elastic tensor (bulk modulus) | 12.511 ± 1.075 | 0.298 ± 0.048 | 16.423 | 0.343 | GPa | RF |
| Exfoliation energy | 52.082 ± 8.855 | 0.938 ± 0.106 | 26.712 | 0.909 | eV/atom | LLM |
| High entropy alloy hardness | 56.163 ± 7.704 | 0.351 ± 0.05 | 81.38 | 0.522 | HV | RF |
| Lithium conductivity | 0.902 ± 0.119 | 0.66 ± 0.073 | 0.928 | 0.912 | S/cm (log scale) | Equal |
| Metallic glass $D_{max}$ | 2.364 ± 0.286 | 0.82 ± 0.31 | 1.948 | 0.842 | mm | LLM |
| Metallic glass $R_c$ | 0.680 ± 0.141 | 0.665 ± 0.126 | 0.779 | 0.658 | K/s (log scale) | Equal |
| Oxide vacancy formation | 1.081 ± 0.037 | 0.405 ± 0.021 | 0.883 | 0.339 | eV | LLM |
| Perovskite formation energy | 0.114 ± 0.003 | 0.273 ± 0.007 | 0.098 | 0.246 | eV/atom | LLM |
| Perovskite O p-band center | 0.146 ± 0.009 | 0.227 ± 0.03 | 0.141 | 0.18 | eV | Equal |
| Perovskite stability | 29.220 ± 1.477 | 0.28 ± 0.028 | 28.73 | 0.236 | meV/atom | Equal |

| Perovskite TEC | 1.469 ± 0.331 | 0.556 ± 0.123 | 1.379 | 0.606 | $K^{-1}$ ($\times 10^{-6}$) | Equal |
|---|---|---|---|---|---|---|
| Perovskite work function | 0.430 ± 0.036 | 0.324 ± 0.029 | 0.55 | 0.512 | eV | RF |
| Phonon frequency | 62.928 ± 6.492 | 0.273 ± 0.059 | 77.036 | 0.319 | $cm^{-1}$ | RF |
| Piezoelectric max displacement | 0.411 ± 0.030 | 0.825 ± 0.06 | 0.294 | 0.603 | $C/m^2$ (log scale) | LLM |
| Steel yield strength | 99.178 ± 8.444 | 0.519 ± 0.114 | 93.976 | 0.689 | MPa | Equal |
| Superconductivity $T_c$ | 0.166 ± 0.004 | 0.356 ± 0.008 | 0.211 | 0.497 | K (natural log scale) | RF |
| Thermal conductivity | 2.962 ± 0.141 | 0.376 ± 0.052 | 2.787 | 0.634 | W/m-K | LLM |
| Thermal expansion | $5.9\times10^{-6}$ ± $4.1\times10^{-7}$ | 0.361 ± 0.067 | $6.85\times10^{-6}$ | 0.319 | $K^{-1}$ | RF |

The comparison of LLaMA 3 vs. random forest performance is also shown graphically in **Figure 4**, where in **Figure 4** the RMSE/$\sigma_Y$ values are plotted for each property to better observe the random forest vs. LLaMA 3 agreement with a unitless error metric. By taking the average RMSE/$\sigma_Y$ value for all properties, we find that random forest averages 0.450 while LLaMA 3 averages 0.489, an 8.6% increase in average error. This is consistent with the fact that, when compared to random forest, LLaMA 3 results are typically worse by a larger margin than they are better, as can be seen in **Figure 4**. It is worth noting that the LLaMA 3 average RMSE/$\sigma_Y$ value is sensitive both to the datasets considered here and the use of a single test-train split, since the RMSE/$\sigma_Y$ value can be sensitive to the particular test data split evaluated. For example, for the thermal conductivity data, LLaMA 3 produced a lower MAE vs. random forest but a higher RMSE/$\sigma_Y$ because the standard deviation of the chosen test data split was much lower than the overall dataset standard deviation. This effect is diminished when performing evaluations over many test data splits, which was not practical at this time considering the large computational cost of fine-tuning each LLaMA 3 model. Overall, when considering model performance based on MAE, we find that fine-tuning LLaMA 3 using only composition strings as input results in a test error compared to random forest that is better, on par, or worse each about a third of the time, but makes larger average errors when it underperforms, given about a 10% average larger MAE. This finding illustrates that LLaMA 3 is useful for performing regression even on small materials datasets, is capable of internally formulating an effective featurization based on minimal input (i.e., just composition strings), and can provide comparable performance to standard machine learning models like random forest, but with no physical input. These results show that LLaMA 3 (and likely other LLMs) have promise as standard regression models useful for materials property prediction. Although they yield similar performance, fine-tuning an LLM is at this point a significantly more computationally expensive task than using random forest (e.g., several hours of GPU time to finetune the LLM vs. just seconds of CPU time to fit random forest). However, in cases where featurization is not possible or very difficult, the featurization-free LLM-based regression may be a method of choice.

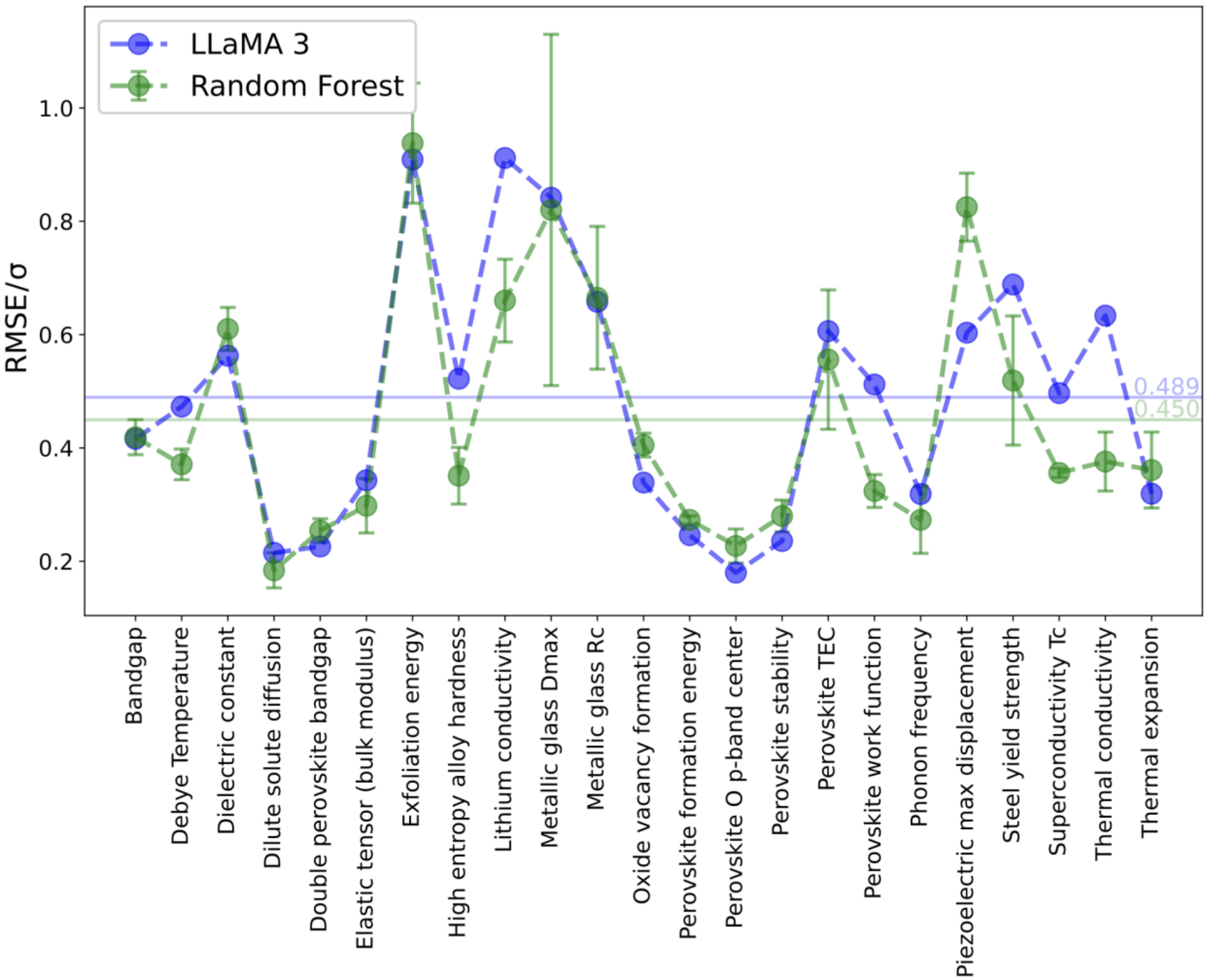

**Figure 4.** Summary of LLaMA 3 and random forest model performance on various materials properties. The plotted values are RMSE/$\sigma_y$ on 20% held out data, split equally into validation and testing in the case of LLaMA 3. The error bars on the random forest values are the standard deviation from 25 splits of random 5-fold cross validation from the work of Jacobs et al.[48] The solid blue and green lines and numbers next to them denote the average RMSE/$\sigma_y$ value across all of the examined materials properties for LLaMA 3 and random forest, respectively.

As an example of a very large dataset of materials property (the 24$^{th}$ materials data set we studied), we fine-tune LLaMA 3 on formation energies in the OQMD database. **Figure 5** provides a learning curve, similar to the above case of QM9 formation energies in **Figure *1***, showing drastic improvement in test MAE as a function of training dataset size. Overall, training on the largest dataset size of 634k for 3 epochs resulted in a test MAE of 0.054 eV. This value is better than a random forest fit using elemental features, which obtained 0.067 eV,[49] is essentially equal to the result from the fully-connected deep neural network ElemNet,[22] which obtained an MAE of 0.055 eV, and is about 2.3× larger than the state-of-the-art model Representation Learning from Stoichiometry (RoosT), which obtained an extremely low MAE of only 0.024 eV. The RoosT result is very impressive considering full atomic structures were not used and seems to be the state of the art for prediction on this dataset without using structural information.[49] Broadly, these results mirror those obtained for the QM9 molecular properties fits and fits to other materials property datasets, where LLaMA 3 can offer regression errors of similar quality to random forest models fit using elemental features, but generally higher errors (by at least a few multiples) compared to state-of-the-art deep neural network approaches.

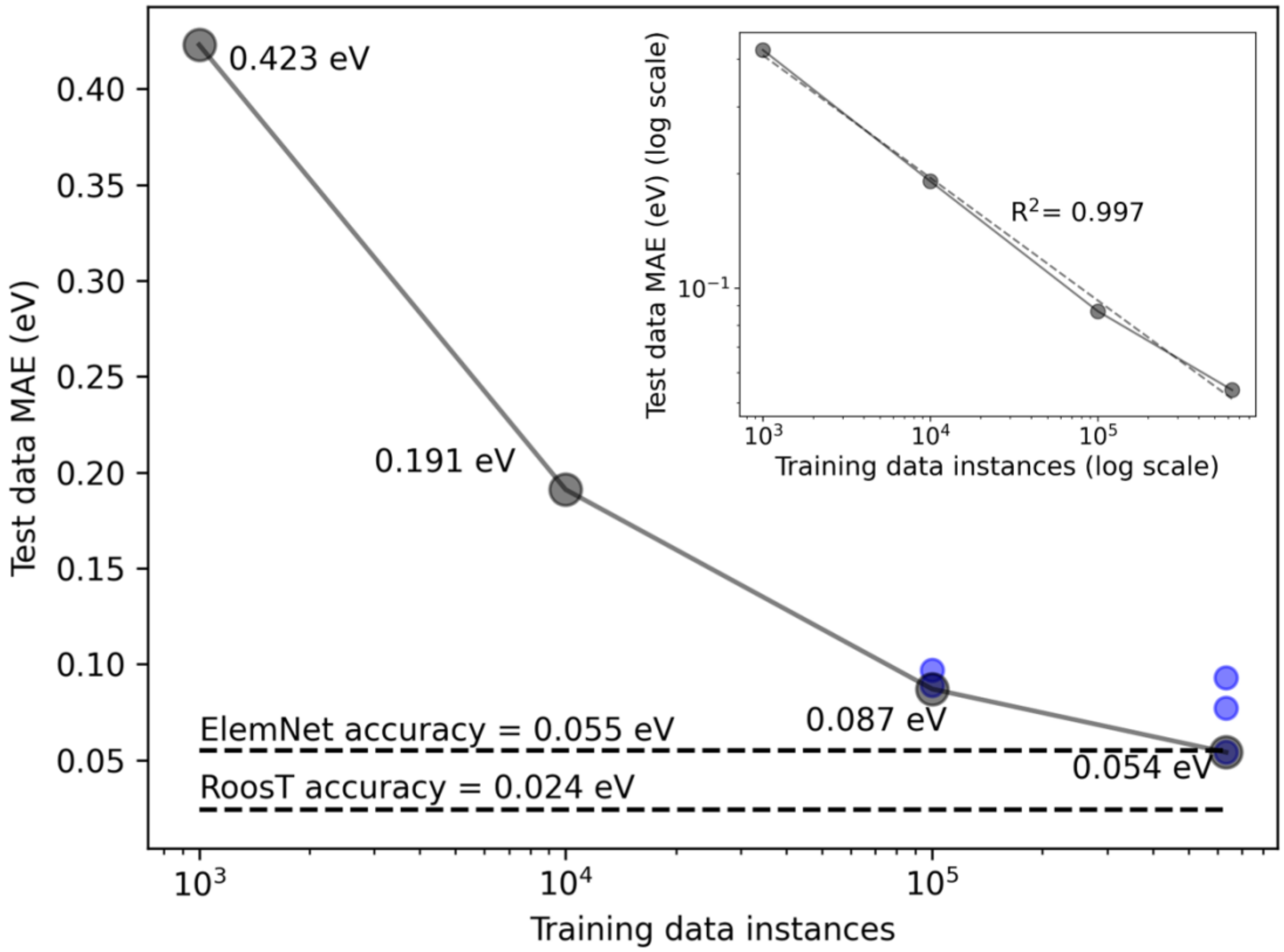

**Figure 5.** Learning curve for LLaMA 3 fine-tuning on OQMD formation energy data. The inset figure depicts the linear trend of log error vs. log number of training points, showing power law scaling. The fit line has a high $R^2$ of 0.997, a slope of -0.322 and intercept of 0.580.

As a final test we compare our fine-tuned LLaMA 3 with AtomGPT (another LLM-based method) by reproducing the results of regression on materials properties performed by Choudhary with AtomGPT.[16] **Table 4** summarizes the results of this test, which show that fine-tuned LLaMA 3 slightly outperforms AtomGPT on two properties, and is slightly outperformed on another two. Fine-tuning an LLM is a very straightforward process that can be applied without modification to almost any LLM, whereas AtomGPT requires using a modified regression head, so the former approach might be preferred for simplicity. On the other hand, AtomGPT uses the LLM primarily as a featurizer and then optimizes the target of interest (RMSE) as a loss function, which one might expect to perform better in most situations. It is also worth noting, that while these tests serve the purpose of comparing the two methods, they do not necessarily represent the actual performance of the model on these properties, due to the quality of the underlying datasets. For example, the superconductivity $T_c$ dataset is a very small dataset of unbalanced data, where just guessing 0K results in better performance (MAE = 1.45 K) than both models as well as all other models tested in Ref. [16]. Similarly, the band gap datasets are very unbalanced and effectively mix regression with classification. They oversimplify the complexity of the band gap property by providing accurate values of energy separation between bands of certain symmetry and orbital composition and symmetry for systems where this value is positive, while indicating 0 eV in cases where this value is zero or would be negative. There are also instances of approximately duplicate entries, which are not strictly duplicate due to slight numerical differences, but are from the physical standpoint, and it is unclear how these impact the model's performance. It therefore remains an open question which is preferable, a direct fine tuning of the LLM generative loss like that described here, or an LLM-based featurization followed by more traditional regression approach like that used in AtomGPT, and more research in this area is needed.

**Table 4.** Summary of LLaMA 3 materials property fits compared to previously published values using AtomGPT from Ref. [16].

| Property name | AtomGPT MAE | Llama 3 MAE |
|---|---|---|
| Eform (eV/atom) | 0.072 | 0.067 |
| OPT Eg (eV) | 0.139 | 0.178 |
| MBJ Eg (eV) | 0.319 | 0.388 |
| Tc (K) | 1.54 | 1.46 |

All of the results presented in this work of fine-tuning LLaMA 3 for molecular and materials property regression point to LLaMA 3 (and likely other LLMs) being a powerful featurization engine, where providing minimal input information in the form of composition and compact-form structure information (e.g., SMILES strings) and materials compositions (only chemical and no structure information) can produce meaningful regression results. We speculate the larger errors of LLaMA 3 relative to state-of-the-art on QM9 properties vs. materials properties is due to the role of detailed structure-based information in the state-of-the-art QM9 model featurization. The state-of-the-art model PAMNet, which formed the basis of the QM9 property comparisons, is a sophisticated deep graph neural network which was able to leverage key features of molecular structure information from quantum mechanics-based simulations. By comparison, LLaMA 3 performed similarly with QM9 models without access to structural information from Pinheiro et al.[45], and essentially on par with random forest models of materials properties trained using elemental features, which contain no structural information. While an initial effort providing structural information to LLaMA 3 (see **Sec. 3.2**) did not provide significantly improved results, more work needs to be done to assess the ability of LLMs like LLaMA 3 to incorporate structural information, such as refining how the information is provided and the fine-tuning is performed.

## 4. Summary and Conclusion:

This work provides an initial examination of the ability of LLMs to perform supervised regression fits to molecular and materials properties. Somewhat surprisingly, we found that the performance of LLMs on regression tasks improves even when solely optimizing for generative loss, which is counterintuitive given that regression tasks typically involve optimizing for mean squared error or similar metrics. We speculate that this ability is due to the featurization and associated distances between compounds induced by generative loss being in some ways similar to those induced by minimizing mean squared error, although more work on this aspect is needed to understand the process better. Through examining fits to more than 25 molecular and materials properties of vastly varying dataset sizes (a couple hundred data points to more than 600k data points), we broadly find that the LLaMA 3 LLM models finetuned using chemical composition and SMILES encodings of molecules can produce regression errors competitive with, although typically somewhat worse than, conventional random forest and fully-connected deep neural network models fit to a suite of elemental features and SMILES encodings, respectively. However, at least at this time and with the fine-tuning procedure used here, LLaMA 3 on QM9 significantly underperforms the state-of-the-art deep neural networks that that have access to detailed structural information about the molecules. Furthermore, the LLaMA 3 fine tuning is much slower than using random forest, and therefore the latter still emerges as preferable for modeling the data explore here.

Interestingly, we find that LLaMA 3 outperforms GPT-3.5 and GPT-4o, suggesting the exact LLM model used may have a large impact on the fit quality, in particular if there are limitations in the choice of fine-tuning hyperparameters. Further, we found that the type of input features used can have a more modest but significant effect on prediction accuracy, where we found that fits to QM9 properties using SMILES vs. InChI strings resulted in a 15-20% error difference, where SMILES always resulted in improved fits. This finding highlights not only the importance of model choice, but also the choice of input used to fine-tune LLMs. Additional questions regarding the best LLM model, best input approach, for example, the use of explicit coordinates, some joint approach leveraging multiple molecular string types, or other string types not investigated here (e.g., SELFIES strings) are worth investigation, as well as the potential impact of prompt engineering and different training approaches, such as leveraging relationships between properties to perform transfer or multitask learning. Overall, this work highlights the versatility of LLMs, suggesting that these models can transcend their traditional applications and address complex physical phenomena, thus paving the way for future research and applications in materials science and other scientific domains.

**Acknowledgements:**

Support for R. J., M. P., L. E. S. and D. M. was provided by the National Science Foundation under NSF Award Number 1931298 “Collaborative Research: Framework: Machine Learning Materials Innovation Infrastructure”. Support for H. M. and V. H. was provided by the National Science Foundation under NSF Award 2020243 “AI Institute: Planning: Institute for AI-Enabled Materials Discovery, Design, and Synthesis”.

**Data and Code Availability:**

All datasets used to fit the LLaMA 3 models, results of all runs, as well as python scripts used to fine-tune the models and perform inference, are available on Figshare:
https://doi.org/10.6084/m9.figshare.26928439.v1 and
https://doi.org/10.6084/m9.figshare.26936770.v1 .

**Conflicts of Interest:**

The authors have no conflicts of interest to declare.